\documentclass[aps,prl,twocolumn,showpacs,showkeys,draft,groupedaddress]{revtex4-1} %,preprint
  \usepackage{color}
  \usepackage{ulem}

\begin{document}

\title{The Wu-Yang potential of Magnetic Skyrmion from SU(2) Flat Connection}

\author{Ji-rong Ren}
\email{renjr@lzu.edu.cn}

\author{Hao Wang}
\email[Corresponding Author: ]{wanghao16@lzu.edu.cn}

\author{Zhi Wang}

\author{Fei Qu}

\affiliation{Institute of Theoretical Physics, Lanzhou University, P.R.China, 730000}

\date{\today}

\begin{abstract}
  The  theoretical  research of the origin of magnetic skyrmion is very interesting.
  By using decomposition theory of gauge potential and the gauge parallel condition of local bases of $su(2)$ Lie algebra, its $SU(2)$ gauge potential is expressed as flat connection.  As an example of application, we obtain  the inner topological structure of second Chern number by $SU(2)$ flat connection method.  It's well known that if magnetic monopole exists in electrodynamics, its Wu-Yang potential is indispensable in $U(1)$ invariant electromagnetic field.  In $2$-dim magnetic materials, we prove that if magnetic skyrmion exists, its integral kernel must be  $U(1)$ Wu-Yang curvature,  where its  $U(1)$ Wu-Yang potential is the projection of $SU(2)$ flat connection on $su(2)$ local Cartan subalgebra.  The magnetic skyrmion can be created by performing concrete $SU(2)$ local gauge transformation to  $su(2)$ Cartan subalgebra $\sigma_3$. The components of the $U(1)$ Wu-Yang curvature correspond to the emergent electromagnetic field of magnetic skyrmion.
 \end{abstract}



\maketitle

Magnetic skyrmions are topological   spin configurations of magnetizations, which usually originate from the chiral  Dzyaloshinskii-Moriya
  interactions  and have been observed in the laboratory\cite{skyrmionobservation,ultrathinmagskyr}. They have stimulated the research interest associated with both spintronics and information storage\cite{Skyrmionsonthetrack2013,logicgate2015} and have also  provided a broad stage for
    gauge field theory. While investigating the topological Hall effect and the Berry phase in magnetic nanostructures\cite{BrunoDugaevTaillefumier}, P.
   Bruno et al. developed an $SU(2)$ gauge transformation technique and obtained an electromagnetic-like vector potential.  In addition, G.
   Tatara obtained an adiabatic spin gauge field  by constructing a type of $SU(2)$ gauge transformation for ferromagnetic metals\cite{Tatara}.
  Because magnetic skyrmions can be locally written or deleted in various spin technologies by varying the magnetization direction\cite{Skyrmionsonthetrack2013,logicgate2015,Nikals1},
   we consider that only the change of the magnetization direction instead of its position indicates that it undergoes local gauge transformation. Thus, the $SU(2)$ gauge transformation of magnetization  are observed to occur naturally rather than having to be introduced.
   In this  letter, we discover the reason because of which the $SU(2)$ gauge connection $A_\mu$ can be expressed in terms of $SU(2)$ gauge group element $U$ by investigating the Duan-Ge gauge potential decomposition theory\cite{DuanGe1979} and  the gauge parallel condition\cite{Witten} of local bases of $su(2)$ Lie algebra.  Further, by defining the components of $U$ as the
     components of a unit vector field and using Duan's topological current theory\cite{topologicalcurrent,Duanzhang}, we will  derive that the second Chern number
    can be expressed
    in terms of the Hopf indices and Brouwer degrees at all zero points of this vector field.  Several physicists who have investigated magnetic monopoles have discussed the $U(1)$ Wu-Yang potential and the invariant electromagnetism tensor in $SU(2)$ gauge field theory \cite{DuanGe1979,Cho1980,FaddeevNiemi1999}. Here, we will
       prove that the Wu-Yang potential is the projection of an  $SU(2)$ flat connection onto the local bases of the $su(2)$ Cartan subalgebra, and the integral kernel of a magnetic
       skyrmion is proportional to the curvature of this Wu-Yang potential,  which is related to the emergent electromagnetic field of magnetic skyrmion.

Let  $P(M,SU(2),\pi)$  be the principal bundle on a four-dimensional orientable compact base manifold $M$.
  The local bases of $su(2)$ Lie algebra are defined\cite{Duanzhang,WalkerMichaelStevenDuplij} as
\begin{equation}\label{1}
  U\sigma_iU^\dagger \equiv u_i\;\;,\;\; i=1,2,3,
\end{equation}
 where $U $ is  an element of the $SU(2)$ group and Pauli matrices $\sigma _i\;(i=1,2,3)$ form the basis of  the $su(2)$ Lie algebra.  $SU(2)$ group elements $U $  can be expressed as
\begin{equation}\label{2}
 U=U^0+iU^a\sigma_a,
\end{equation}
and their components satisfy
\begin{equation}\label{normalization}
 (U^0)^2+U^aU^a=1\;,\;\;\; a=1,2,3.
\end{equation}
It can be easily proved that $u_iu_j+u_ju_i=2\delta_{ij}$.

  Each local basis $u_i$ is still  Lie algebra vector and can be spanned by
\begin{equation}\label{3}
u_i=u^a_i\sigma_a,
\end{equation}
where $u^a_i=[(U^0)^2-(U^bU^b)]\delta^a_i-2U^0U^b\epsilon_{bia}+2U^aU^i$.

The covariant derivatives of $u_i$ can be written in one-form as
\begin{equation}\label{5}
Du_i=du_i-ig[A,u_i]
\end{equation}
where A is the $SU(2)$ gauge potential one-form.

According to Duan's gauge potential decomposition and inner structure theory\cite{DuanGe1979},  the $SU(2)$ gauge potential can be decomposed into  two parts as
\begin{equation}\label{6}
A=a+b,
\end{equation}
where $a$ satisfies the gauge transformation
\begin{equation}\label{7}
a'=UaU^{-1}+\frac{i}{g}dUU^{-1},
\end{equation}
and $b$ satisfies the adjoint transformation
\begin{equation}\label{8}
b'=UbU^{-1}.
\end{equation}

Next, we define the Clifford scalar product as
\begin{equation}\label{10}
u_i\cdot u_j=\frac{1}{2}(u_iu_j+u_ju_i)=\delta_{ij}
\end{equation}
By this definition,  we obtain $u_iu_i=\delta_{ii}\equiv3$ .

Further, it is well known that the gauge potential $A$ is an $su(2)$ Lie algebra vector,  and it  can also be decomposed in terms of the local basis of the $su(2)$ Lie algebra
\begin{equation}\label{11}
A=(A\cdot u_j)u_j\;,\;\;(j=1,2,3).
\end{equation}
By substituting eq.(\ref{11}) into eq.(\ref{5}), we obtain
\begin{equation}\label{12}
Du_i=du_i-ig(A\cdot u_j)[u_j,u_i].
\end{equation}
Because  $\left[u_j,u_i\right]=2u_ju_i-2\delta_{ij}$, this becomes
\begin{equation}\label{14}
Du_i=du_i-2ig(A\cdot u_j)u_ju_i+2ig(A\cdot u_i).
\end{equation}
Multiplying eq.(\ref{14}) with $u_i$ from the right to obtain
\begin{eqnarray}\label{15}
(Du_i)u_i&=&(du_i)u_i-6ig(A\cdot u_j)u_j+2ig(A\cdot u_i)u_i \nonumber  \\
&=&(du_i)u_i-4igA .
\end{eqnarray}
Thus, the $SU(2)$ gauge potential $A$ can be rewritten as
\begin{equation}\label{16}
A=\frac{1}{4ig}(du_i)u_i-\frac{1}{4ig}(Du_i)u_i.
\end{equation}

By the definition  $u_i\equiv U\sigma_iU^\dagger$ and the unitarity of the gauge transformation, we  can  directly prove that
 \begin{eqnarray} \label{17}
    du_i     &=&  \left( d U\right)U^\dagger   U \sigma_iU^\dagger     +U\sigma_i U^\dagger U  \left(d U^\dagger  \right)   \nonumber  \\
    &=&  [dUU^\dagger,u_i].
 \end{eqnarray}
Multiplying eq.(\ref{17}) with $u_i$ from the right yields
\begin{equation}\label{18}
 du_iu_i=3dUU^\dagger-u_idUU^\dagger u_i.
\end{equation}
Because
\begin{eqnarray}\label{19}
dUU^\dagger&=&(dUU^\dagger\cdot u_i)u_i \nonumber \\
&=&\frac{1}{2}(dUU^\dagger u_i+u_idUU^\dagger)u_i \nonumber \\
&=&\frac{3}{2}dUU^\dagger+\frac{1}{2}u_idUU^\dagger u_i,
\end{eqnarray}
we obtain $ u_idUU^\dagger u_i=-dUU^\dagger $.
Substituting  this into eq.(\ref{18}) gives as $ du_iu_i=4dUU^\dagger $;hence, the gauge potential $A$  can be decomposed as
 $A=\frac{1}{ig}dUU^\dagger-\frac{1}{4ig}(Du_i)u_i $.
By considering the gauge parallel condition\cite{Witten} $Du_i=0$, we obtain the flat  gauge potential expressed in terms of gauge transformation $U$ as
\begin{equation}\label{flatconnection}
A=\frac{1}{ig}dUU^\dagger .
\end{equation}
The following discussion will denote that this flat connection method can help in revealing  the underlying relations between the topological charges (such as Chern number, monopole and skyrmion etc.) and the gauge transformations. In particular, the gauge transformations of magnetization can write or delete magnetic skyrmions on 2D magnetic materials.

As an application of  the $SU(2)$ flat connection, now we investigate the second Chern number. Specifically, we will use Duan's topological current theory\cite{topologicalcurrent} to reveal the inner structure of the second Chern number, which is expressed in terms of the Hopf indices and Brouwer  degrees at all the zero points.

It is extensively known\cite{Chern} that the second Chern form can be expressed as
\begin{equation}\label{30}
C_2(P)=d\Omega,
\end{equation}
where $
\Omega=\frac{g^2}{8\pi^2}Tr(A\wedge dA-\frac{2}{3}igA\wedge A\wedge A)$
is the Chern-Simons $3$-$form$.
 By substituting eq.(\ref{flatconnection}) into $\Omega$ and by applying unitary condition $UU^\dagger=1$, we obtain
\begin{equation}
\Omega= \frac{1}{24\pi^2}Tr(dUU^\dagger\wedge dU\wedge dU^\dagger).
\end{equation}
Further, the second Chern form can be given as
\begin{equation}\label{32}
C_2(P)=-\frac{1}{24\pi^2}Tr(dU\wedge dU^\dagger\wedge dU\wedge dU^\dagger).
\end{equation}
The second Chern class $C_2$ is the integral of its Chern form $C_2(P)$ over the base manifold $M$ that can be given as
\begin{equation}
C_2=\int_M C_2(P).
\end{equation}
Using eq.(\ref{2}), we can express $U$ as
\begin{equation}\label{33}
U=U^0+iU^a\sigma_a=U^As_A \;\;,\;\;(a=1,2,3; A=0,1,2,3),
\end{equation}
where $s_A=(I,i\vec{\sigma})$.
Thus,  eq.(\ref{32}) can be rewritten as
\begin{eqnarray}\label{SecondChernForm}
C_2(P)
&=&-\frac{1}{24\pi^2}\epsilon^{\mu\nu\lambda\rho}\partial_\mu U^A\partial_\nu U^B\partial_\lambda U^C\partial_\rho U^D Tr(s_As_B^\dagger s_C s_D^\dagger)d^4 x  \nonumber  \\
&=&-\frac{1}{12\pi^2}\epsilon^{\mu\nu\lambda\rho}\epsilon_{ABCD}\partial_\mu U^A\partial_\nu U^B\partial_\lambda U^C\partial_\rho U^D d^4 x
\end{eqnarray}
By considering the unitary condition eq.(\ref{normalization}), we can introduce a four-component field $\phi^A(x)\;(A=0,1,2,3)$  and define $U^A=\frac{\phi^A}{\|\phi\|}$,
yielding
\begin{equation}\label{differentialofUA}
dU^A=\frac{d\phi^A}{\|\phi\|}+\phi^Ad(\frac{1}{\|\phi\|}).
\end{equation}
By substituting eq.(\ref{differentialofUA}) into eq.(\ref{SecondChernForm}), we can denote that
\begin{equation}
C_2(P)=-\frac{1}{4\pi^2}\frac{\partial^2}{\partial\phi^A\partial\phi^A}(\frac{1}{\|\phi\|})J(\frac{\phi}{x})d^4x,
\end{equation}
where $J(\frac{\phi}{x})$ is the Jacobian and $\epsilon^{ABCD}J(\frac{\phi}{x})=\epsilon^{\mu\nu\lambda\rho}\partial_\mu\phi^A\partial_\nu\phi^B\partial_\lambda\phi^C\partial_\rho\phi^D$.
Further, we know that the Green's function formula in $\phi$-space satisfies$
\frac{\partial^2}{\partial\phi^A\partial\phi^A}(\frac{1}{\|\phi\|})=-4\pi^2\delta^4(\phi)$,
therefore
\begin{equation}\label{deltachernclass}
C_2(P)=\delta^4(\phi)J(\frac{\phi}{x})d^4x.
\end{equation}
According to Duan¡¯s theorem\cite{topologicalcurrent},  if $\phi^A(x)\;(A=0,1,2,3)$ contains $\ell$ isolated zeros $x=z_i\;(i=1,2,...,\ell)$,  we will obtain
\begin{equation}
\delta^4(\phi)=\sum_{i=1}^{\ell}\frac{\beta_i\delta^4(x-z_i)}{|J(\frac{\phi}{x})|_{x=z_i}},
\end{equation}
where $\beta_i$ is the Hopf index of the $\imath$th zero. Given the definition of the Brouwer degree $
\eta_i=\frac{J(\frac{\phi}{x})}{|J(\frac{\phi}{x})|}=sgn[J(\frac{\phi}{x})]|_{x=z_i}=\pm1
 $,
 the second Chern form eq.(\ref{deltachernclass}) can be formulated as
\begin{equation}
C_2(P)=\sum_{i=1}^{\ell}\eta_i\beta_i\delta^4(x-z_i)d^4x,
\end{equation}
hence, the second Chern number $C_2$ is
\begin{equation}
C_2=\int_M C_2(P)=\int_M \sum_{i=1}^{m}\eta_i\beta_i\delta^4(x-z_i)d^4x=\sum_{i=1}^{m}\eta_i\beta_i.
\end{equation}

For a given base manifold $M$, $C_2$ is observed to be a topological invariant.
 From eq.(\ref{32}), we can conclude that  the topological charge, i.e. second Chern number, is a gauge invariant for arbitrary local $SU(2)$ gauge transformation.
 Direct application of the flat connection method and the $SU(2)$ group elements method had enabled us to  observe that the second Chern  number  is left unchanged by the $SU(2)$ gauge transformation,which gives us some  insights into  the  relation between the magnetic skyrmions and $SU(2)$ group elements.

While studying magnetic monopoles,  't Hooft\cite{Hooft1974}, Duan-Ge\cite{DuanGe1979},
Cho\cite{Cho1980}, and Faddeev and Niemi\cite{FaddeevNiemi1999}
  observed that a non-Abelian $SU(2)$ gauge field theory with an $SU(2)$ electromagnetic field tensor is required to describe the
  magnetic monopoles, with all but 't Hooft observing that the decomposition theory of $SU(2)$ gauge potential played an important role in this description.
   Duan and Zhang\cite{Duanzhang} summarized these  results and extended them to the $SU(N)$ case. The corresponding $SU(N)$ invariant electromagnetic
   tensor  can be given as
\begin{equation}\label{33}
f^i_{\mu\nu}=(F_{\mu\nu},n_i)+\frac{i}{g}(n_i,[D_\mu n_k, D_\nu n_k]),\;\; i=1,2,...,N-1,
\end{equation}
where
\begin{equation}\label{LocalBasCar}
n_i \equiv UH_iU^\dagger,\;\;\;i=1,...,N-1
\end{equation}
  forms the  Abelian and local bases of the $SU(N)$ Cartan subalgebra  proposed by Prof. Duan\cite{DuanGe1979,Duanzhang, WalkerMichaelStevenDuplij}; further,  $H_i$'s  also forms the basis of the  Cartan subalgebra of $SU(N)$ Lie algebra.  $H_i$ commute  with each other, and it is easy to show that $n_i$  also  commutes with each other. Finally,$F_{\mu \nu }$ denotes the $SU(N)$ gauge curvature tensor as
 \begin{equation}\label{SU(N)GauCurvTens}
F_{\mu\nu}=\partial_\mu A_\nu-\partial_\nu A_\mu-ig[A_\mu, A_\nu].
 \end{equation}
 The $SU(N)$ gauge potential $A_\mu$ can be projected onto the direction parallel to the $n_i$ and orthogonal to  $n_i\;$\cite{Duanzhang}
 \begin{equation}\label{AmuProj}
 A_\mu=A_\mu^kn_k+[[A_\mu,n_k],n_k],
 \end{equation}
 and we can also prove that
 \begin{equation}\label{CommofAmuPro}
   \left([[A_\mu,n_k],[A_\nu,n_k]],n_i\right)=-([A_\mu,A_\nu],n_i).
 \end{equation}
 By substituting eq.(\ref{AmuProj}) into eq.(\ref{SU(N)GauCurvTens})  using eq.(\ref{CommofAmuPro}), we can obtain $(N-1)$  invariant electromagnetic tensors
 \begin{equation}\label{36}
 f_{\mu \nu}^i=\partial_\mu A_\nu^i-\partial_\nu A_\mu^i+K^i_{\mu\nu},
 \end{equation}
 where
  \begin{equation}\label{Amunu}
K^i_{\mu\nu}=\frac{i}{g}(n_i,[\partial_\mu n_k,\partial_\nu n_k]) ,
 \end{equation}

 From the definition of the local  bases of Cartan subalgebra in eq.(\ref{LocalBasCar}),we can easily denote that
   \begin{equation}\label{PartialOfLocaBas}
 \partial_\mu n_i=[\partial_\mu UU^\dagger, n_i].
  \end{equation}
  Further, eq.(\ref{PartialOfLocaBas}) can be straightforwardly rewritten in a covariant derivative form as, i.e.
   $
 \partial_\mu n_i-ig[A_\mu, n_i]=0$,    where $
  A_\mu=\frac{1}{ig}\partial_\mu UU^\dagger
 $
  is exactly the flat connection as that presented in eq.(\ref{flatconnection}).

 By substituting eq.(\ref{PartialOfLocaBas}) into eq.(\ref{Amunu}), we can denote that $K_{\mu\nu}$ is proved as the curvature tensor of the Wu-Yang potential
\begin{equation}\label{Wu-YangCurvature}
K^i_{\mu\nu}=\partial_\mu a^i_\nu-\partial_\nu a^i_\mu,
\end{equation}
where
\begin{equation}\label{Wu-YangPotential}
 a^i_\mu=(A_\mu,n_i) \;\;(i=1,...,N-1)
 \end{equation}
  is the $i$-th U(1) Wu-Yang potential. This is the gauge potential of the magnetic monopole in $SU(2)$ case\cite{Wu-Yang1,DuanGe1979};however,
 in this study, it is exactly
the Abelian projection of the flat connection $A=\frac{1}{ig}dUU^\dagger$ in the $n_i$ direction.

In the $SU(2)$ gauge theory, there is only one Cartan subalgebra basis element $\sigma_3$. We have defined its corresponding  local basis element as $n_3=U\sigma_3 U^\dagger$ which  is essentially $u_3$. In the context of magnetic skyrmions
in 2D magnetic materials, $\sigma_3$ can be viewed as the Pauli matrix along $z$-direction; therefore,
the local Cartan basis $n_3$ can be naturally treated as  the unit magnetization $\vec{m}$ of the magnetic material.
$$
 n_3 =U\sigma_3 U^\dagger \equiv  \vec{m}.
$$
 This equation connects the magnetization with the local basis of Cartan  subalgebra through  gauge transformation.
Thus,  the curvature tensor of the Wu-Yang potential $K^i_{\mu\nu}$   in eq.(\ref{Amunu}) becomes
\begin{equation}\label{k3munu}
K^3_{\mu\nu}=\frac{i}{g}(\vec{m},[\partial_\mu \vec{m},\partial_\nu \vec{m}]).
\end{equation}

 Magnetic skyrmions are a type of quasi-particle that have a vortex-like spin configuration and carry a characteristic topological charge $S$ and are observed in magnetic materials. If $\Sigma$ is a 2D manifold of 2D magnetic
material thin film, we can observe  that the  skyrmion charge in this film is \cite{RenYu2017}
  \begin{equation}\label{skyrmion}
   S=\frac{1}{8\pi}\int_{\Sigma }\vec{m}\cdot (\frac{\partial \vec{m}}{\partial x^{\mu}} \times  \frac{\partial \vec{m}}{\partial x^{\nu}})dx^{\mu}\wedge dx^{\nu}\;
  (\mu,\nu=1,2),
  \end{equation}
 where $x^1$ and $x^2$ denote the coordinates of the manifold $\Sigma$.

By comparing eq.(\ref{k3munu}) with eq.(\ref{skyrmion}), we find that the curvature tensor $K^3_{\mu\nu}$ of the $SU(2)$ Wu-Yang potential  is proportional to the skyrmion's integral kernel if a surface with coordinates $\{x^\mu;\;\mu =1,2\}$ in the 2D manifold $M$ is selected . When one needs to create a $Bloch$ type or $N\acute{e}el$ type magnetic skyrmions in 2D magnetic materials, we can orient the magnetization in the desired direction by performing local $SU(2)$ gauge transformations to $\sigma _3$ at all the points. In addition, the emergent local electromagnetic fields $B^e_i$ and $E^e_i$ can be written as\cite{Schulz} $\;
 B^e_i(x,y)= \frac{\hbar}{2}\epsilon_{ijk}\vec{m}\cdot(\partial_j\vec{m}\times\partial_k\vec{m}),
 E^e_i(x,y)= \hbar\vec{m}\cdot(\partial_i\vec{m} \times\partial_t\vec{m})
 $.
By comparing these with eq.(\ref{skyrmion})(\ref{k3munu}), we can conclude that the emergent electromagnetic field is proportional to the
components of the Wu-Yang curvature tensor. Further, the topological Hall effect of magnetic skyrmions in 2D magnetic materials  is essentially  the scattering of the electrons by this emergent electromagnetic field.

By analyzing the invariant electromagnetic tensor that was  proposed and studied by Wu, Yang, 't Hooft, Duan, Ge, Cho, Faddeev and Niemi while investigating the magnetic monopoles and their electrodynamics, the Wu-Yang curvature has emerged as a part of the invariant electromagnetic tensor. Thus, we  can conclude  that  the Wu-Yang potentials of magnetic monopoles will exist in $U(1)$ invariant electromagnetic fields if the magnetic monopoles exist in electrodynamics. Our observation that the Wu-Yang curvature is proportional to the magnetic skyrmion's integral kernel is considerably important. Thus, the existence of magnetic skyrmions as topological charges in 2D magnetic materials, implies the presence of a corresponding Wu-Yang potential. Further, magnetic skyrmions can be created by performing concrete local gauge transformations to  $su(2)$ Cartan subalgebra $\sigma_3$, demonstrating their theoretical origins  in $SU(2)$ gauge field theory. This theory suggests a method for the creation of  magnetic skyrmions in the laboratory.
In our future work, we will focus on the relationship between the Wu-Yang potential and the Berry connection\cite{BrunoDugaevTaillefumier}.

\end{document}